\documentclass[aps,prl,twocolumn,showpacs,superscriptaddress]{revtex4}
\usepackage{graphicx}
\usepackage{psfrag}
\usepackage{ae}
\usepackage{amsmath,amssymb}

\begin{document}
\date{\today}

\title{Ions at the air-water interface: An end to one hundred year old mystery?}

\author{Yan Levin}
\affiliation{Instituto de F\'{\i}sica, Universidade Federal do Rio Grande do Sul, Caixa Postal 15051, CEP 91501-970, 
Porto Alegre, RS, Brazil}
\author{Alexandre P. dos Santos}
\affiliation{Instituto de F\'{\i}sica, Universidade Federal do Rio Grande do Sul, Caixa Postal 15051, CEP 91501-970,
Porto Alegre, RS, Brazil}
\author{Alexandre Diehl}
\affiliation{Departamento de F\'{\i}sica, Instituto de F\'{\i}sica e Matem\'atica,
Universidade Federal de Pelotas, Caixa Postal 354, CEP 96010-900, Pelotas, RS, Brazil}

\begin{abstract}

Availability of highly reactive halogen ions at the surface of aerosols has tremendous implications for the
atmospheric chemistry. Yet neither
simulations, experiments, nor existing theories  are able to provide a fully consistent 
description of the electrolyte-air interface.  In this paper a new theory is proposed which allows us to
explicitly calculate the ionic density profiles, the surface tension,  and the
electrostatic potential difference across the solution-air interface. Predictions of the theory are compared to 
experiments and are found to be in excellent agreement.  
The theory also sheds new light on
one of the oldest puzzles of physical chemistry  --- the Hofmeister effect.  

\end{abstract}


\maketitle

Since van't Hoff's experimental measurements of osmotic pressure more than $120$ 
years ago,
electrolyte solutions have fascinated physicists,
chemists, and biologists alike~\cite{Le02}. The theory of Debye and Hückel (DH)~\cite{DeHu23} was able to
address almost all of the properties of bulk electrolytes.  On the other hand,
electrolyte-air interface remains a puzzle up to now. The mystery appeared when 
Heydweiller~\cite{He10} measured the surface tension of various electrolyte solutions and observed that
it was larger than the interfacial tension of pure water.  While the dependence on 
the type of cation was  weak, a strong variation of the excess surface
tension was found with the type of anion.  
The sequence was 
reverse of the famous Hofmeister series~\cite{Ho88}, which was known to govern 
stability of protein solutions against salting-out. An explanation for this behavior
was advanced by Wagner~\cite{Wa24} and Onsager and Samaras~\cite{OnSa34} (WOS), who argued that when 
ions approach the dielectric air-water interface, 
they see their image charge and are repelled from it.  
This produces a depletion zone which, with the help of thermodynamics, can be related to the
excess surface tension. The theory and its future modifications~\cite{Le00}, however, were  
unable to account for the Hofmeister series and
showed  strong  deviations  from the 
experimental measurements above $100$mM concentrations. The fact that something  was seriously
wrong with the WOS approach was already clear in $1924$, when Frumkin measured 
the potential difference across the air-water interface and found that for  all halogen salts 
 --- except for fluoride ---  the electrostatic potential difference (air $-$ water)  was more negative for
solution  than for pure water~\cite{Fr24}.  This
suggested that anions were able to approach the interface closer than the cations, or even be adsorbed to it! This
contradicted the very foundation of the WOS theory.  The confused state of affairs continued for the 
next $70$ years, until the  photoelectron emission 
experiments~\cite{MaGi91,Ga04,Gh05} and the polarizable force fields simulations~\cite{DaSm93}
showed that Frumkin was right, and ions might be present at the interface.  The situation, however, remains far from resolved.
Simulations predict so much adsorption that the excess surface tension of NaI solution becomes
negative, contrary to experiments~\cite{IsMo07}.  Furthermore, while the electron
spectroscopy was finding the surface composition of
solution to be enhanced in anions~\cite{Gh05}, vibrational sum-frequency spectroscopy (VSFS) indicated a significantly
diminished anion population in the topmost layer~\cite{RaRi04}.
The two results {\it appear} to be contradictory. 
The questions, therefore, remain: Are there ions at the air-water interface?
If so, why are they  there and what are their concentrations?  Besides its relation to the Hofmeister series,
availability of highly reactive halogens at the surface of aerosol particles has  a tremendous implication
for the atmospheric chemistry and might help to explain the excessive rate of ozone depletion 
observed experimentally~\cite{KnLa00}.  In this Letter a theory will be presented 
which allows all the pieces of this  hundred year old 
puzzle to fit together.

We begin by studying the  excess surface tension $\gamma$ of an electrolyte solution. This 
can be calculated by integrating the Gibbs 
adsorption isotherm equation, $ {\rm d} \gamma=-\Gamma_+ {\rm d} \mu_+ - \Gamma_- {\rm d} \mu_-$, where 
$\Gamma_\pm$ are the ion excess per unit area, and $\mu_\pm$ are the ionic chemical potentials.  
Suppose that the electrolyte is
confined to a mesoscopic drop of water of radius $R$,  corresponding to the position of the Gibbs 
dividing surface (GDS)~\cite{HoTs03}.  Adsorptions are defined as
\begin{equation}
\label{e1}
\Gamma_\pm \equiv \frac{1}{4 \pi R^2} \left[\int_0^\infty \rho_\pm(r) 4 \pi r^2 {\rm d}r -\frac{4 \pi R^3}{3} c_b \right] \,,
\end{equation}
where $\rho_\pm(r)$ are the ionic density profiles and 
$c_b=\rho_+(0)=\rho_-(0)$ is the bulk concentration of electrolyte. 
If the system --- water$+$vapor --- contains $N$ ion pairs,  
Eq. (\ref{e1}) simplifies to 
$\Gamma\pm=N/4\pi R^2 - c_b R/3$.   

Let us first consider the alkali metal cations such as Li$^+$, Na$^+$, and K$^+$. These ions  
are small and strongly hydrated.   They can, therefore, be modeled as rigid spheres of {\it hydrated} 
radius $a_h$ and fixed charge $q$ located at the center.  
Water and air will be treated as uniform dielectrics of
permittivities $\epsilon_w=80$ and $\epsilon_a=1$, respectively,   with a discontinuity across the 
GDS.  For a cation to move across the GDS,  requires shedding its hydration sheath.  
For small, highly hydrated cations, this costs a lot of energy, resulting  
in a high potential energy barrier and a strong hardcore-like repulsion from the GDS.

Suppose that a cation is located at position $\bf r_p$ from the center of the drop.  
The electrostatic potential inside the electrolyte satisfies the usual 
DH equation,  $\nabla^2 \varphi-\kappa^2 \varphi=-\frac{4 \pi q}{\epsilon_w} \delta({\bf r-r_p})$, 
where $\kappa=\sqrt{8 \pi q^2 c_b/\epsilon_w k_B T}$ is the inverse Debye length.  
In the vapor phase the
electrostatic potential satisfies the Laplace equation,
$\nabla^2 \varphi=0$. For mesoscopic water drops of radius $R \gg 1/\kappa$, curvature 
can be neglected and 
the two partial differential equations can be solved using the  Hankel transform~\cite{LeMe01}.  
Once the electrostatic potential is known, the work required to bring an ion 
from the bulk electrolyte to some distance $z$ from 
the interface --- $z$ axis is oriented into the drop, with the GDS at $z=0$ ---
can be calculated using the Güntelberg charging process~\cite{Gu26}.  
We find,
\begin{eqnarray}
\label{e2}
&&W(z;a_h)=  \\
&&\frac{q^2}{2\epsilon_w} \int_0^\infty dk e^{-2 s (z-a_h)} 
\frac{k[ s \cosh(k a_h)-k \sinh(k a_h)]}
{s[ s \cosh(k a_h)+ k \sinh(k a_h)]} \nonumber\;,
\end{eqnarray}
where $s=\sqrt{\kappa^2 + k^2}$.  Eq.~(\ref{e2}) accounts for two fundamental contributions: 
the interaction of an ion with its image across the interface,  and for the loss of  screening  
resulting from breaking of spherical symmetry near the surface.   
In their theory of surface tension,
Onsager and Samaras included  ionic size by 
adopting the bulk spherically symmetrical  potential of Debye and Hückel.  
By doing this,  they have failed to account for the loss of screening
near the interface, which leads to additional repulsion.  This is one of the reasons why WOS 
theory significantly underestimates surface tensions of ``hard'' 
non-polarizable electrolytes such as NaF~\cite{LeMe01}.

While the alkali metal ions are strongly hydrated, the large halogen anions, such as iodine and 
bromide, have low electronic charge density and are weakly hydrated.  
To solvate an ion of radius $a_0$, requires creation of a cavity and  disturbance of the hydrogen bond network. 
For small cavities of radius $a_0<4$ \AA$,$   
the free energy cost scales with the volume of the void~\cite{LuCh99}.  
If part of the ion leaves the aqueous environment, the cavitational 
energy will decrease proportionately to the volume exposed.  This results
in a short range cavitation potential which forces ions to move 
across the air-water interface,
\begin{eqnarray}
\label{e3}
U_{cav}(z)=\left\{
\begin{array}{l}
 \nu a_0^3 \,; z \ge  a_0  \\
 \frac{1}{4} \nu a_0^3  \left(\frac{z}{a_0}+1\right)^2 \left(2-\frac{z}{a_0}\right); -a_0<z<a_0.
\end{array}
\right.
\end{eqnarray}
From bulk simulations~\cite{RaTr05}, we obtain  $\nu \approx 0.3 k_B T/$\AA$^3$. 
For strongly hydrated alkali metal cations, the cavitational energy cost is too low to compensate the loss
of hydration and the exposure of ionic charge to the low dielectric environment.  
For soft, unhydrated halogens,  the situation is very different.
As these ions move across the interface, they progressively shift their charge towards the part that remains
hydrated, thus allowing them to reduce the cavitational energy 
at a small price in electrostatic self energy~\cite{Le09}. 

To see how this works, consider a perfectly polarizable ion  modeled as 
a conducting spherical shell of radius $a_0$ and charge $q$, free to distribute itself over its  surface. 
The electrostatic self
energy ~\cite{Le09} of such an ion when its center is located at distance $z$ from the GDS is
$U_s(z)=\frac{q^2}{2 \epsilon_w a_0}\frac{1} {\frac{\arccos(z/a_0)}{\pi} +\frac{\epsilon_a}{2 \epsilon_w}}$.
This expression is accurate for $-a_0<z<a_0/4$, and is exact for ions located precisely at the GDS, $z=0$.  
For such ions, $U_s(0) \approx q^2/\epsilon_w a_0$, 
which is more than an order of magnitude lower than the electrostatic energy of a
hard non-polarizable ion at the same position, $\sim q^2/4\epsilon_a a_0$!

Although fundamentally important at the interface, 
for distances $z \ge a_0$, effects of ionic polarizability are negligible. 
This can 
be  verified by noting that for a hard ion located at $z=a_0$, the electrostatic self 
energy is $U_{hard}(a_0)= 3 q^2/4 \epsilon_w a_0$.    
On the other hand, the self energy of a perfectly polarizable ion at the same position is  
$U_{soft}(a_0)=q^2/2 \ln(2)\epsilon_w a_0$, which can be calculated exactly by
resumming a series of images 
necessary to keep the ion at fixed potential.  For $a_0 \approx 2$\AA, the difference between $U_{hard}(a_0)$ and $U_{soft}(a_0)$
is about $0.1 k_B T$. Therefore, for distances $z>a_0$, 
the ionic polarizability can be neglected.  
The above calculation was performed in the infinite dilution limit.  At finite ionic concentrations, 
polarization effects will be even less significant, since all the induced interactions 
are doubly screened~\cite{FiLeLi94}.
For $z>a_0$ the anion-interface electrostatic potential will, therefore, be well approximated  
by Eq.~(\ref{e2}), with $a_h \rightarrow a_0$.
The total
anion potential can then be obtained by interpolating between Eq.~(\ref{e2}) and $U_s(z)$. We find
\begin{eqnarray}
\label{e5}
U_{tot}(z)=\left\{
\begin{array}{l}
W(z;a_0)+\nu a_0^3+\frac{q^2}{2 \epsilon_w a_0} \, \text{ for } z \ge  a_0  \\
\frac{W(a_0;a_0) z}{a_0} + U_s(z)+U_{cav}(z)\, \text{ for } 0<z<a_0 \,. \\
U_s(z)+U_{cav}(z)\, \text{ for } -a_0<z \le 0
\end{array}
\right.
\end{eqnarray}
Since the electrostatic self energy is extremely large for $z<-a_0$, no ion will be found at these distances.  

\begin{figure}
\begin{center}
\includegraphics[width=6cm]{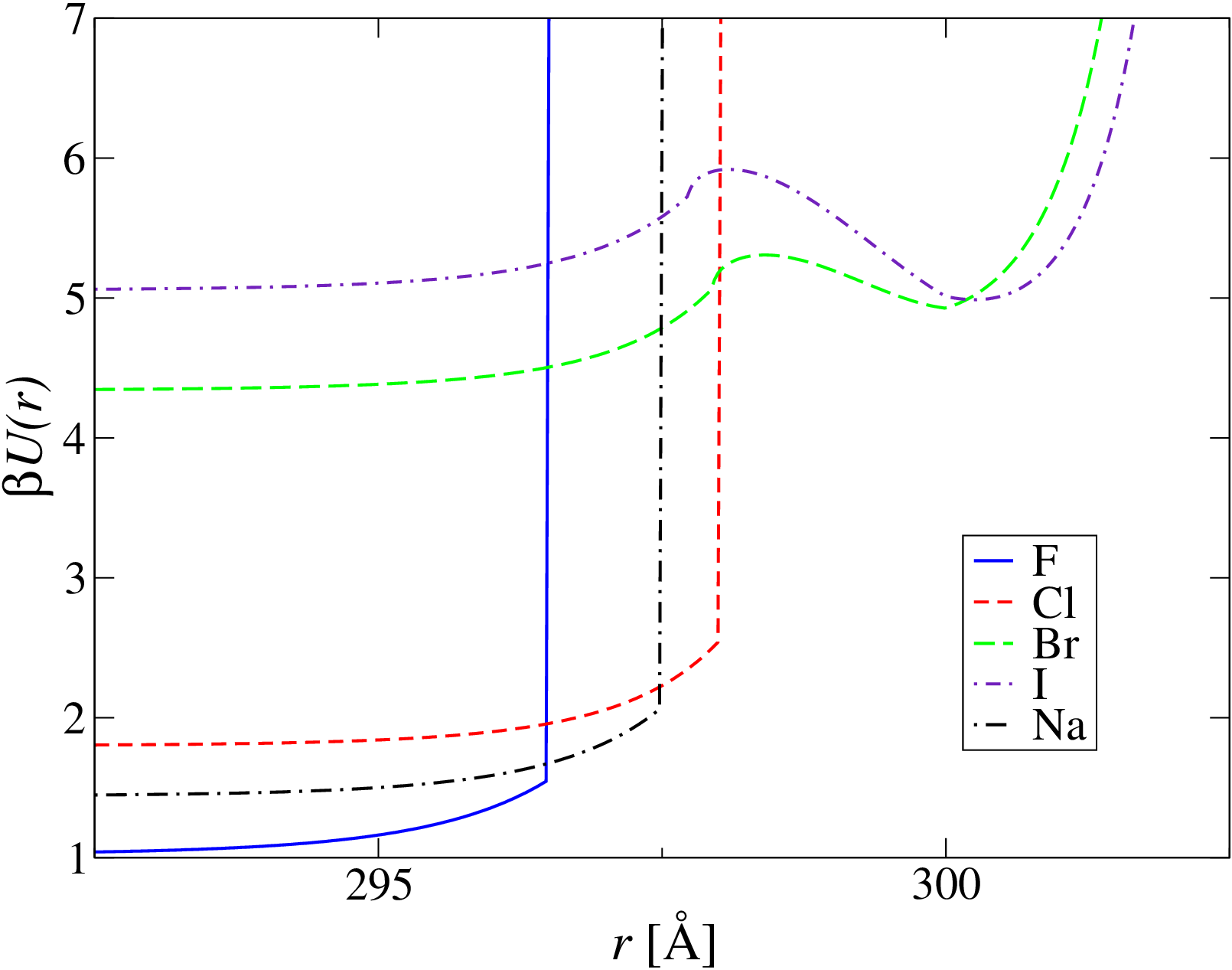}
\end{center}
\caption{The ion-surface interaction potentials at 1M salt concentration.  For hydrated ions there is a hardcore repulsion
from the GDS located at $r=300$\AA, and the cavitational energy is omitted since it is constant.
}
\label{fig0}
\end{figure}

So far our discussion has been for perfectly polarizable ions.  
Real ions, however, have finite polarizability.  The polarization potential for such ions has been 
derived in reference~\cite{Le09}.
For such ions, Eq.~(\ref{e5}) should  be modified by replacing the ideal potential $U_s(z)$, by the polarization
potential $U_p(z)$ of reference~\cite{Le09}. The potentials of all ions are plotted in Fig. \ref{fig0}.

The ionic density profiles can now  be calculated numerically by solving the non-linear modified 
Poisson-Boltzmann equation (mPB) 
$\nabla^2 \phi(r)=-\frac{4\pi q }{\epsilon_w} \left[\rho_+(r)-\rho_-(r)\right]$, where 
\begin{eqnarray}
\label{e7}
\rho_-(r)&=& \frac{N e^{\beta q \phi(r) -\beta U_{tot}(r)}}
{\int_0^{R+a_0} 4\pi r^2\, dr \, e^{\beta q \phi(r) -\beta U_{tot}(r)}}  \\
\rho_+(r)&=& \frac {N \Theta (R-a_h-r) e^{-\beta q \phi(r)-\beta W(z;a_h)}}{\int_0^{R-a_h} 4\pi r^2\, dr\, e^{-\beta q \phi(r)-\beta W(z;a_h)}} \nonumber \;,
\end{eqnarray}
and $\Theta$ is the Heaviside step function.
The excess surface tension of electrolyte solution can then 
be obtained by integrating the Gibbs adsorption isotherm equation (\ref{e1}) with $\beta \mu_\pm=\ln(c_b \Lambda_\pm^3)$,
where $\Lambda_\pm$ is the de Broglie thermal wavelength.    

\begin{figure}
\begin{center}
\includegraphics[width=6cm]{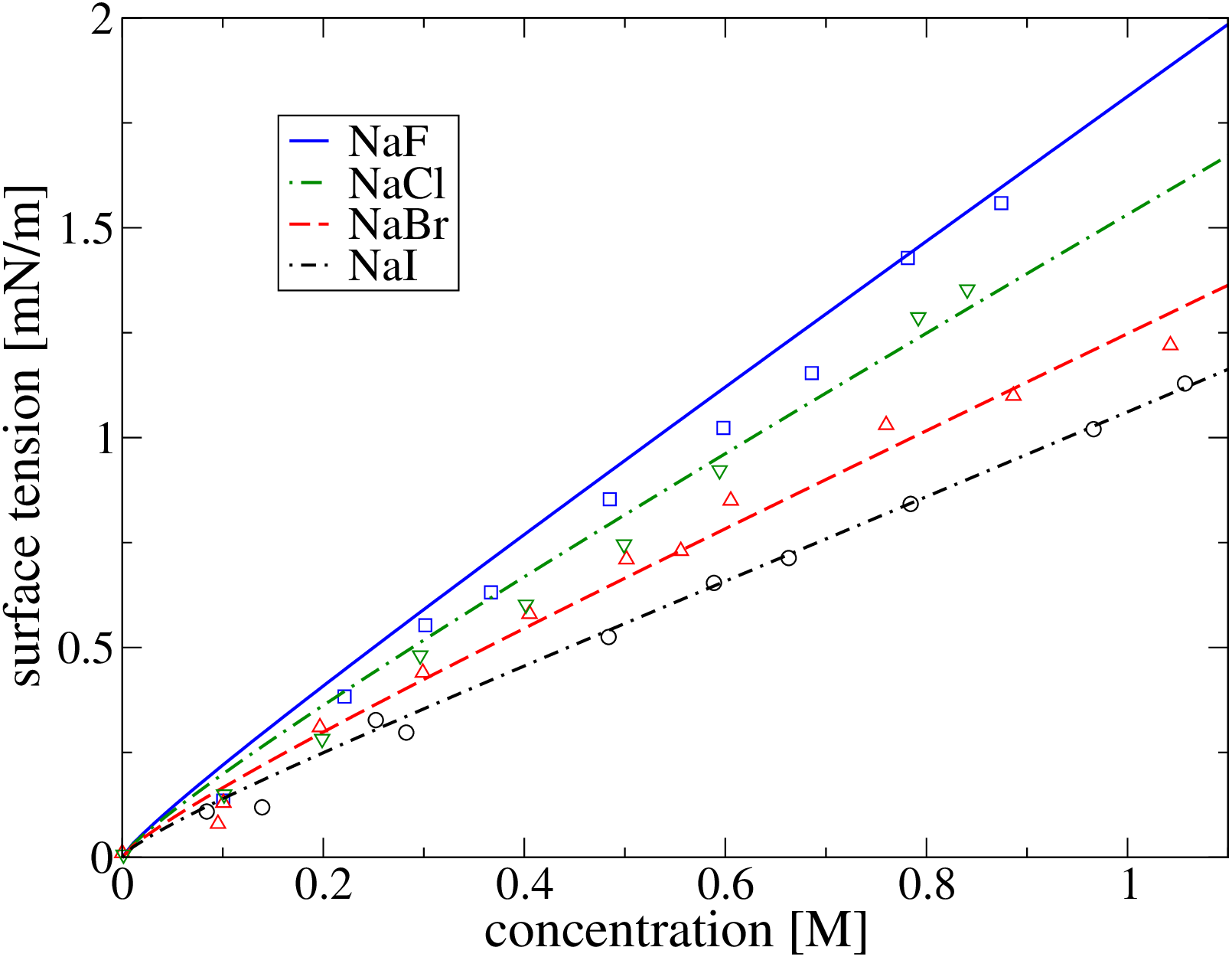}
\end{center}
\caption{Surface tension of NaF, NaCl, NaBr, and NaI.  Na$^+$ and Cl$^-$  are 
partially hydrated with  $a_{h}=2.5$ \AA$\,$ and
$2.0$ \AA$\,$, respectively, 
F$^-$ is fully hydrated with $a_h=3.52$ \AA, while the large halogens I$^-$ and  Br$^-$ are unhydrated. 
Symbols are the experimental data from \cite{MaTs01}.
}
\label{fig1}
\end{figure}

We start with NaI.  Since I$^-$ is 
large and soft, it should be unhydrated in the interfacial region. For its radius, 
we  use the value calculated by 
Latimer, Pitzer, and Slansky~\cite{LaPi38} from fitting the experimentally measured free 
energy of hydration to the 
Born model.  Latimer radii for halogens come out to be 
almost identical to the Pauling crystal radii, differing from them by only 0.1 \AA. 
Since in the bulk our theory reduces  to the Born model,  Latimer 
radii:  $a_I=2.26$ \AA,  $a_{Br}=2.05$ \AA,  $a_{Cl}=1.91$ \AA, and  $a_{F}=1.46$ \AA, 
are particularly 
appropriate. For ionic 
polarizabilities we will use the values from reference~\cite{PyPi92}:
$\gamma_I=7.4$ \AA$^3$, $\gamma_{Br}=5.07$ \AA$^3$, $\gamma_{Cl}=3.77$ \AA$^3$, and  $\gamma_{F}=1.31$ \AA$^3$.  
Our strategy will be to 
adjust the hydrated radius of Na$^+$ to best fit the experimental surface tension for NaI.   
The same value 
of $a_h$, will then be used to calculate the surface tension of other sodium salts and compare them 
to the experimental measurements~\cite{MaTs01}.
 Fig. \ref{fig1} shows the result of this procedure.						
We find that $a_{h}=2.5$ \AA$\,$ gives an excellent fit to 
the experimental data for NaI.  Since Br$^-$ ion is also large and soft,
we expect that it will also remain unhydrated  in the interfacial region.  
This expectation is well justified, and a good agreement is obtained with the
experimental data, Fig. \ref{fig1}. The situation should be
very different for F$^-$, which is small, hard, and  strongly hydrated. 
This means that just like for cation, a hard core repulsion from the 
GDS must be explicitly included in the mPB equation.
We then find an almost perfect agreement with the experimental data using the bulk hydrated 
radius of  F$^-$, $a_h=3.52$ \AA$\,$ \cite{Ni59}.  
The most difficult electrolyte to study theoretically is the usual table salt, NaCl.  The size of chloride is 
sufficiently small that hydration  must be taken into account. At the same time, it is 
also quite polarizable.  We find that using Latimer size for Cl$^-$ gives a very reasonable agreement with experiment, however,
the agreement can be  made perfect if we assume a small hydration, $a_h=2.0$ \AA.

\begin{figure}
\begin{center}
\includegraphics[width=8cm]{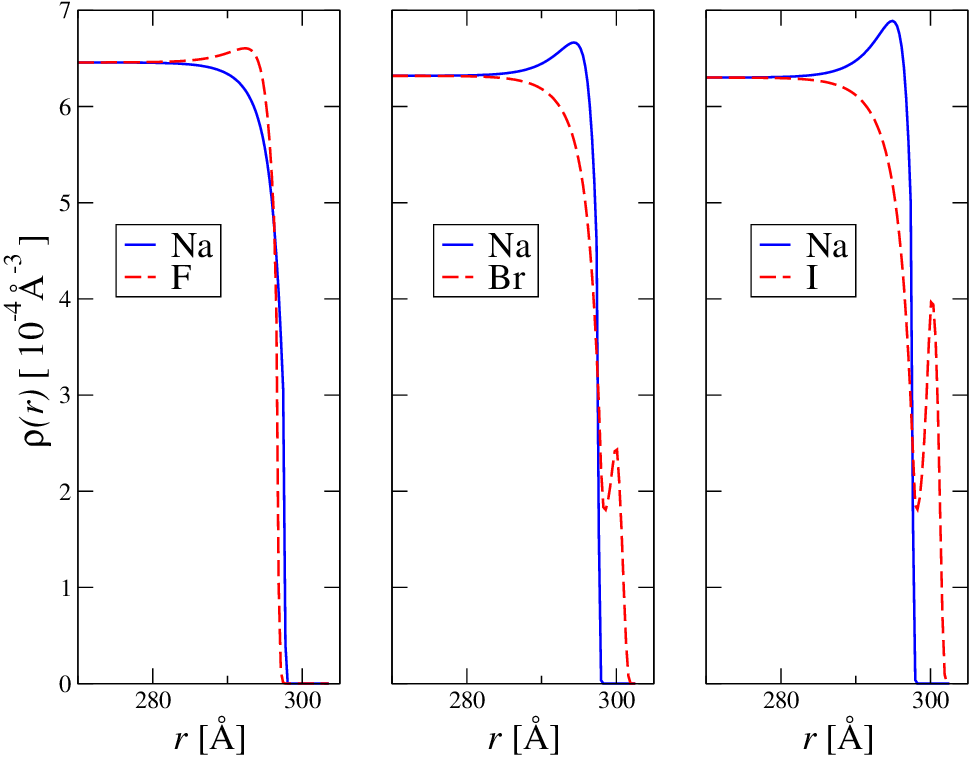}
\end{center}
\caption{Density profiles for NaF, NaBr, and  NaI at 1M concentration. The GDS is at $r=300$\AA.}
\label{fig2}
\end{figure}

The density profiles for NaF, NaBr, and NaI are shown in Fig.\ref{fig2}.
In agreement with the polarizable force fields simulations, 
the density profiles for large halogens I$^-$ and Br$^-$  differ significantly from the classical WOS picture.
We find that there is a considerable concentration of anions at the GDS. 
However, unlike simulations~\cite{IsMo07},  and in agreement with the
surface tension experiments, our  adsorption always remains negative.
The theory also agrees with the electron spectroscopy measurements showing that close to the  
GDS, there is a larger excess of 
anions {\it over} cations.  However, just as was found using VSFS,
the absolute concentration of anions at the surface is about
half that of the bulk.
We are, therefore, able to reconcile the two sets of {\it apparently} contradictory experimental results.
Finally, we calculate the excess electrostatic potential difference across the 
interface, $\Delta \chi$, for 1M solutions of NaF, NaCl, NaBr, and NaI .  We obtain: 
$+3.8$, $-1.9$, $-9.1$, and $-14.0$ mV, for the four salts respectively. These are quite
close to the values originally measured by Frumkin~\cite{Fr24,Ra63}.   In particular, one should note 
the change of sign of the 
electrostatic potential going from NaF to NaCl, 
first observed by Frumkin and confirmed by the present theory.

We have seen that by adjusting {\it only} the hydrated radius of sodium cation, we are able to account for 
the surface tensions of
four different electrolyte solutions and for their values of $\Delta \chi$.  
It should, therefore, be possible to predict the
surface tensions and the $\Delta \chi$ of other salts  --- as long as their  anions are sufficiently
large and weakly hydrated.  This is the case, for example, for 
NaNO$_3$, NaIO$_3$, and  NaClO$_4$. The only experimental data available to us is 
for NaNO$_3$ \cite{MaYo07} which, once again, 
shows a good agreement between the theory and experiment. For NaClO$_4$, we find that at low concentrations
the excess surface tension is very small (slightly negative) but grows with increasing concentration of electrolyte. 
For 1M NaClO$_4$, the calculated value of $\Delta \chi$ is $-42$ mV, while the
value originally  measured by Frumkin was $-48$ mV ~\cite{Ra63}.  For 1M  
NaNO$_3$ and NaIO$_3$, we obtain $\Delta \chi=-8.2$ and $-22$ mV, respectively.

We have presented a theory which allows us to calculate surface tensions and surface potentials 
for seven different electrolyte solution using only {\it one}
adjustable parameter  --- the hydrated radius of sodium cation, $a_h=2.5$ \AA. 
This value is very reasonable, lying between 
the Pauling crystal radius and the bulk
hydration radius of Na$^+$. In the case of five sodium salts for which there is  experimental data available to us, 
the theory is found to be in very good agreement with the measured surface tensions. 
Using the {\it same} value of $a_{h}$, we are also able to account
for the experimentally measured  electrostatic potential differences across the solution-air interfaces. 
The theory  provides a very interesting  picture of the interfacial region: 
alkali metal cations and fluoride anion are strongly hydrated and are repelled 
from the GDS.  On the other hand, heavy halogens, 
Br$^-$ and I$^-$, and the monovalent oxy-anions,  NO$_3^-$, IO$_3^-$, and  ClO$_4^-$, are 
unhydrated, and as a result of their polarizability are significantly adsorbed to the surface.   Nevertheless, their absolute
concentration at the GDS remains below that of the bulk value.  All these conclusions are  
in agreement with recent photoelectron spectroscopy and the VSFS measurements. 
In view of the success of the theory, it seems reasonable to hope that a fully 
quantitative understanding of the 
Hofmeister effect might now be in sight.

YL would like to thank C.J. Mundy for comments on the manuscript.
This work is partially supported 
by CNPq, INCT-FCx, and by the US-AFOSR under the grant FA9550-06-1-0345.


\end{document}